## Title

- Electron acceleration in direct laser-solid interactions far beyond the ponderomotive limit.

## Authors

Meng Wen,[1,2]* Yousef I. Salamin,[1,3] Christoph H. Keitel[1]

## Affiliations

[1]Max-Planck-Institute for Nuclear Physics, Saupfercheckweg 1, 69117 Heidelberg, Germany.
[2]Department of Physics, Hubei University, Wuhan 430062, China.
[3]Department of Physics, American University of Sharjah, POB 26666, Sharjah, United Arab Emirates

*Corresponding author. Email: meng.wen@mpi-hd.mpg.de

## Abstract

In laser-solid interactions, electrons may be generated and subsequently accelerated to energies of the order-of-magnitude of the ponderomotive limit, with the underlying process dominated by direct laser acceleration. Breaking this limit, realized here by a radially-polarized laser pulse incident upon a wire target, can be associated with several novel effects. Three-dimensional Particle-In-Cell simulations show a relativistic intense laser pulse can extract electrons from the wire and inject them into the accelerating field. Anti-dephasing, resulting from collective plasma effects, are shown here to enhance the accelerated electron energy by two orders of magnitude compared to the ponderomotive limit. It is demonstrated that ultra-short radially polarized pulses produce super-ponderomotive electrons more efficiently than pulses of the linear and circular polarization varieties.

## Introduction

Generation of energetic electrons by laser interaction with matter has witnessed considerable development over the past four decades. In the interaction of a laser pulse with an under-dense plasma (ambient electron density $n_e \leq 0.03 n_c$, where $n_c$ is the critical density) (*1*) GeV-energy electron beams can be obtained via wakefield acceleration (*2*). Larger numbers of electrons can be accelerated in denser plasmas ($n_e > 0.1 n_c$) where direct laser acceleration is dominant, to energies below the ponderomotive limit (*3-6*). The term *ponderomotive limit* refers to the maximum energy gain, $mc^2 \left( a_0^2/2 - 1 \right)$, by a free electron during interaction with a plane-wave laser pulse (*7, 8*). Here, $a_0 = eE_0/(mc\omega)$ is a dimensionless intensity parameter, in which $E_0$ is the electric field amplitude and $\omega$ its frequency, while $m$ and $-e$ are the mass and charge, respectively, of the electron, and $c$ is the speed of light in vacuum. The ponderomotive scaling (*3-8*) is widely quoted within the context of discussions of direct laser acceleration (*9, 10*) in the interaction of intense laser fields with dense matter ($n_e > n_c$), including solid-density plasmas. Generation of electrons which gain over 7 times the ponderomotive energy in near-critical plasmas ($n_e \lesssim n_c$) has recently been reported (*11-18*). This process is associated with important effects, such as the enhancement of ion acceleration in the near-critical plasmas (*19*).

Electron energy beyond the ponderomotive limit is achieved by anti-dephasing acceleration (ADA). Electrons reach such energies in near-critical plasmas by self-injection into the acceleration phase of the laser field, aided by stochastic motion (*11, 12*), transverse electric fields (*13*), and longitudinal electric (*14, 15*) or magnetic fields (*16, 17*). In the absence of anti-dephasing, energies of the electrons generated in laser-solid interactions hover, generally, around the ponderomotive limit (*3-6*). Only MeV electrons are generated when linearly-polarized (LP) and circularly-polarized (CP) terawatt laser pulses are used. However, an experiment in which a micro-wire, used as an advanced solid target to generate and transport hot electrons over millimeters (*20-22*), has recently demonstrated reaching several times the ponderomotive energy, when LP laser pulses are used (*23, 24*). Near-critical plasmas can be generated in the vicinity of solid targets for some applications (*13, 15*). Here, ADA is put forward as a new mechanism for electron acceleration directly from a solid target. It promises to deliver significantly higher electron energies, and to simplify experimental implementation. Previous work on applications associated with electron acceleration to ponderomotive energies, from laser-solid interactions, include ion acceleration (*25*), fast ignition (*26*), laser hole-boring (*27*), high-order harmonic generation (*28*), half-cycle XUV pulse generation (*29*), Bremsstrahlung x-ray generation (*30*), and generation of terahertz radiation (*31*). Progress in these applications, stands to be advanced by availability of more energetic, shorter and denser electron bunches.

In this article, the novel mechanism of ADA, in the interaction of radially-polarized (RP) laser pulses of ultrashort duration with solid wires, is conceived, intuitively explained and backed up by particle-in-cell (PIC) simulations. Our investigations demonstrate unprecedented attosecond picocoulomb electron bunches with energies around one hundred MeV, i.e., two orders of magnitude higher than the ponderomotive limit. In particular, employing RP pulses, electrons are extracted from a wire target, and accelerated by the laser fields, see Fig. 1 for a schematic. Self-injection with a small dephasing rate is caused by the collective motion of the plasma electrons and the complex laser field variations.

**Results**

**Mechanism of anti-dephasing acceleration (ADA)**

Direct laser acceleration by LP pulses leads to the generation of periodic electron bunches (*32*) and CP pulses generate spiral currents (*22*). In these cases, the azimuthal electron motion (*17*) makes it difficult to define an acceleration phase. By contrast, the acceleration phase in an RP laser field is determined directly by the regions of negative axial electric fields, with their troughs at phases $(2N-1/2)\pi$, where $N$ is an integer (*33*). The radial and axial electric field components ($E_r$ and $E_z$) of the unperturbed laser pulse are shown schematically in Fig. 1(A). When the RP laser pulse propagates along the wire of radius smaller than its own waist radius at focus, electrons get knocked out and subsequently accelerated by the laser field. In each laser cycle, distortion to the electron distribution, and the process of electron injection, can typically be described as follows, with the help of Fig. 1. The force associated with $E_r$ pulls electrons into the vacuum around phase1, and pushes them backwards relative to the target over phase 3. Meanwhile electrons get accelerated by $E_z$ around phase 2 of a half cycle, and decelerated during interaction with the subsequent half cycle. Due to the acceleration around the trough of $E_z$, dephasing around 3 is weaker than that about phase 1. As shown schematically in Fig. 1(B) the perturbed electrons are gathered around, and move in phase 2 with the troughs of $E_z$ in an annular region, see also Fig. 2. As a collective plasma effect, the formation of annular bunches has two advantages. On the one hand, as the laser pulse propagates forward, parts of the gathered electrons dephase to phase 3 and get pushed back to the wire by positive $E_r$. During dephasing, the affected electrons repel and help those of the annular bunch to stay in phase 2. On the other hand, new electrons from phase1 get pre-accelerated by those of the annular bunch. Thus, they are launched in the accelerating phase with reduced dephasing, and injection by ADA is triggered.

**Collective effects of the perturbed electrons**

We consider an ultrashort RP laser pulse with $a_0 = 2$ irradiating a wire target. The laser peak arrives at $z = 0$ at $t = 0$, see also Methods section. Snapshots at $t = 1.5\tau$ (where $\tau = \lambda/c$ is the laser period), of the radial and axial electric fields and perturbation of the electron density $\Delta n_e = n_e - n_0$, are shown in Fig. 2. Over the phase values $\phi \in [\phi_0, \phi_0 + \pi]$, the radial electric field $E_r$ is negative, so it works to knock electrons out of the target wire and the negative $E_z$ acts to accelerate them forward, while the opposite happens over the interval $[\phi_0 - \pi, \phi_0]$. Here $\phi_0 = -\pi/2$ is the initial phase of the laser pulse. Figure 2(A) shows the positive $E_r$ around the wire is strengthened, due to perturbation of the electrons, while negative $E_r$ is weakened. Figure 2(C) shows many electrons are collected beyond the cylindrical boundary of the wire, around the phase $\phi_0$, at which the unperturbed trough of $E_z = -|E_{z0}| = \sqrt{2\exp(1)}\lambda E_0/(\pi w_0)$ is expected (34). The negative electric field gradient, induced by electrons in the annular bunch, shifts the trough of $E_z$ to $\sim \phi_0 + \pi/2$ as shown in Fig. 2(B), where the trough of $E_r$ also sits. The near-wire and off-axis $E_z$ are shown by solid-red and dashed-red curves in Fig. 2(D). While the off-axis $E_z$ is unperturbed, the near-wire $E_z$ is determined by the time-averaged field, as a contribution from the space-charge of the perturbed electrons.

Pre-acceleration of the electrons in the combined fields around the wire is enhanced by the large surface-to-volume ratio of the wire target, as illustrated in Fig. 3, where snapshots of the accelerating fields are shown, respectively, at times $t = 0.5\tau$, $0.75\tau$ and $\tau$ in Fig. 3(A)-(C). Trajectories of four electrons are shown injected for later acceleration into different phases. Phase I is centered on $-\pi/2$, and phase II is around $-5\pi/2$. Solid and dashed curves are trajectories of different particles injected into the same phase, and the diamonds in Fig. 3(A)-(C) mark their positions at the corresponding times. Zooming in on the trajectories in Fig. 3(D) reveals that these electrons originate at the front (left) end of the target wire. Prior to entering their respective acceleration phases, the electrons undergo radial and axial oscillations inside the target, forced by the laser field. Subsequently, i.e., when the field structure with forward-shifted negative $E_z$ and weakened negative $E_r$ are acting and the right phase is reached, they get pulled out of the target slowly and accelerated forward with velocities approaching $c$. Figures 3(E) and (F) show evolution in time of their phases and scaled energies, with the gray shades representing the unperturbed acceleration phase. It is shown, for example, that particles following these typical trajectories can stay in the region with negative $E_z$, as they slowly drift in weakened negative $E_r$, and can be injected into the accelerating phase with an injection (pre-acceleration) energy of $\gamma_0 > 1$.

**Energy gain beyond the ponderomotive limit**

ADA occurs after the pre-accelerated electrons are injected into the laser fields. The dephasing rate is given by $R = -\gamma d\phi/(\omega dt)$. Using the relativistic equations of motion, one gets

$$\frac{d\gamma}{d\phi} = \frac{p_z}{R} \frac{eE_z}{m^2 c^2 \omega}, \tag{1}$$

with $p_z$ the axial electron momentum component, and

$$R = \gamma - \frac{p_z}{mc} = 1 - \frac{e}{mc\omega} \int E_z d\phi. \tag{2}$$

The message of Eq. (2) is quite simple: for an electron initially at rest $R=1$, but subjecting the electron subsequently to the action of a symmetrically oscillating $E_z$ alone will cause $R$ to exhibit symmetric oscillations. However, the laser $E_z$, combined with the $E_z$ generated by the plasma electrons [especially when comparable to $E_0/a_0$ (*14*)] will reduce the dephasing rate significantly.

Figures 3(G) and (H) pertain to the electrons during the main acceleration stage. The electrons are launched into that stage by pre-acceleration with small dephasing rates, and get accelerated during a time $\sim \tau/R$. This time duration may be reduced due to the fact that the phase velocity $v_p = c(\pi w_0^4 + \lambda^2 z^2)/(\pi w_0^4 + \lambda^2 z^2 - \lambda^2 w_0^2/2)$ is clearly greater than $c$ for tightly focused lasers. Electrons launched into phase II achieve higher energy gains, because the target becomes hotter with time, allowing more electrons to be knocked out and injected. In this regime, stronger space-charge and pinch effects (*22*) enhance acceleration by the on-axis axial component $E_z$ and boost the pre-acceleration to smaller $R$ (*13*). The dephasing rates shown in Fig. 3(H) hover around $R \sim 0.01$ and can be as small as 0.005. The energy gain receives a boost from the reduction in $R$ via Eq. (1) and approaches $\sim 2\pi e |E_{z0}|/(Rmc\omega)$. Unfortunately, the optimal anti-dephasing conditions, which lead to maximum acceleration, cannot be sustained by all electrons. For example, the dephasing rates of electrons accelerated in phase I do not decrease monotonically, as shown in Fig. 3(H) due to the fact that they move radially off-axis, into regions for which $r > w_0/\sqrt{2} > r_0$, i.e., beyond position of the peak of $E_r$. By contrast, electrons in phase II sustain small $R$ values and acquire higher energy, as shown in Fig. 3(G).

**Discussion**

The self-injection regime of electrons for ADA has been discussed through examples involving RP pulses in interaction with wire targets. The RP laser pulse with a discrete accelerating phase, as illustrated schematically in Fig. 1, generates an electron bunch with FWHM of about 481 attosecond, as shown in Fig. 4(A). More discrete attosecond electron bunches with super-ponderomotive energy can be generated with longer driving pulses. Energy spectra, of electrons driven by laser pulses of amplitudes $a_0 = 2$ and higher, are shown in Fig. 4(B), which shows evolution of the beam energy and charge with the driving laser intensity. Further PIC simulations show that the cutoff energies depend on $r_0$ only weakly, and that the results converge as the resolution is increased. Although the accelerating phase is hard to define in LP and CP pulses, electrons are also gathered at azimuthally-dependent phases. Therefore, the ADA regime works also with LP and CP pulses, resulting in the electrons getting accelerated by a continuous stable phase. 3D-PIC simulations, similar to the above, have been carried out, in which the RP pulses are replaced by LP and CP pulses of the same amplitudes $a_0$, focal radii $w_0$, and temporal envelopes $\eta$ [see Methods, and note that such intense single-cycle laser pulses have been demonstrated experimentally (*35*)]. Electrons accelerated by pulses of all three polarizations have continuous spectra. The cutoff energies ($48mc^2$, $64mc^2$, and $174mc^2$, driven by LP, CP, and RP pulses, respectively) are much higher than the ponderomotive energy gain of $mc^2(a_0^2/2 - 1)$, with $a_0 = 2$. An RP pulse is evidently superior to the CP and LP pulses for generating high-energy electrons. In the RP case, electrons can be launched into discrete phases, as opposed to being injected into continuous phases using the other polarization modes. Thus, the beam charge in the former is smaller than in the latter. Total charges of the electron beams driven by ultrashort pulses, shown in Fig. 4, are of the order of tens of pC, and increase with laser duration (*22, 23*). The ratio of the beam charge to the laser duration is similar to that in a laser-wire experiment of approximately the same intensity (*36*).

Divergences of the electron beams driven by RP, CP and LP lasers, calculated from the momentum distribution of particles $\sigma = \sqrt{\langle (p_x^2 + p_y^2)/p_z^2 \rangle}$, are 28.8, 113.1 and 100.6 mrad, respectively. LP and CP laser pulses generate electron beams of lengths that can be controlled by the pulse duration. While the beam charge decreases with increasing initial density, as shown by the dashed-red line in Fig. 4(D), the energy gain does not get affected drastically, at least for CP pulses.

**Methods**

In our calculations, the wire is assumed to be a cylinder, of length $L$ and radius $r_0$, lying with its axis along $+z$, with $z \in [\lambda, \lambda + L]$. Fields of the incident RP pulse are derived from the normalized vector potential, $e\vec{A}/mc = a\vec{e}_z$, where $a$ is given by (*37*)

$$a = \frac{\eta\sqrt{2}a_0}{\varepsilon\sqrt{1+\alpha^2}} \exp\left[\frac{1}{2} - \frac{r^2}{w_0^2}\frac{1-i\alpha}{1+\alpha^2} - i\tan^{-1}\alpha + i\phi\right], \qquad (3)$$

$\vec{e}_z$ is a unit vector in the direction of $+z$, and $\eta = \operatorname{sech}[(z/c - t)/T]$ replaces the *sinc* function [encountered in (*36*)] as a temporal pulse envelope, in which $2\ln(1+\sqrt{2})T$ is the FWHM of the corresponding intensity profile. In terms of the amplitudes $E_{r0}$ and $B_{\theta 0}$ of the radial electric and azimuthal magnetic field components, the intensity parameter in Eq. (3) is $a_0 = eE_{r0}/(mc\omega) = eB_{\theta 0}/(m\omega)$. Also, $\varepsilon = w_0/z_r$, where $w_0$ is the waist radius, and $z_r = \pi w_0^2/\lambda$ is the Rayleigh length. Furthermore, $\lambda$ is a central wavelength, $\alpha = (z+ct)/(2z_r)$, and $\phi = \phi_0 + \omega(z/c - t)$ is the on-axis phase, with $\phi_0$ a constant.

3D-PIC simulations have been conducted, employing the code EPOCH (*38*), in which a wire target is located in the $20\lambda \times 40\lambda \times 40\lambda$ simulation box (moving window with a boundary which allows for the transmission of both particles and electromagnetic waves) represented by a Cartesian grid of $800 \times 800 \times 800$ cells with 200 (5) micro-particles for electrons (protons) per cell. The initial target radius, length, and density are given, respectively, by $r_0 = 1.5\lambda$, $L = 10\lambda$, and $n_0 = 20n_c$. Target wires of flexible sizes can be made from a number of materials (*20, 23*). An ultrashort pulse, with $a_0 = 2$, $w_0 = 7\lambda$, $\lambda = 800\text{nm}$, $\phi_0 = -\pi/2$, and a duration $T = \tau$ is focused onto the left boundary of the simulation box (at $z = 0$).

**Acknowledgments**

**Funding:** MW is supported by the Chutian Scholars Program in Hubei Province of China. YIS acknowledges support for this work from an American University of Sharjah Faculty Research Grant (AS1802). **Author contributions:** M.W. proposed the project. Y.I.S. performed the analytical derivations. The PIC simulations were carried out by M.W. All authors contributed to the interpretation of the results and the writing of the paper. **Competing interests:** The authors declare no competing financial interests. **Data and materials availability:** Simulation data and figures are accessible from M.W. upon reasonable requests.

**Figures and Tables**

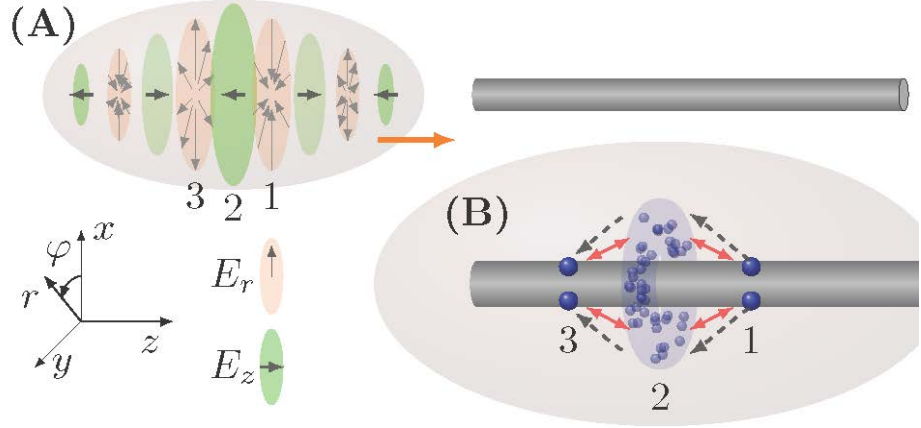

**Fig. 1. Schematic of electron generation from a solid wire target, driven by an RP laser pulse.** (**A**) The pulse, with only its oscillating $E_r$ and $E_z$ shown, is incident upon the left end of the target wire. Troughs of $E_r$ and $E_z$, and an instantaneous peak of $E_r$ are shown at phases 1, 2 and 3, respectively. (**B**) As the pulse propagates along the wire, electrons are typically pulled out of the target, by the negative $E_r$ in phase 1, and subsequently enter phase 2, where they get accelerated by the negative $E_z$. Periodic annular electron bunches are formed at troughs of $E_z$, losing some electrons to phase 3 and getting fed by others from phase 1. Electrons which dephase into phase 3 get pushed back towards the wire by positive $E_r$. Dashed-black arrows show trajectories of electrons in the moving frame of the laser pulse, and red arrows represent the repulsion between electrons in different phases (which yields the reduction in dephasing described in the main text).

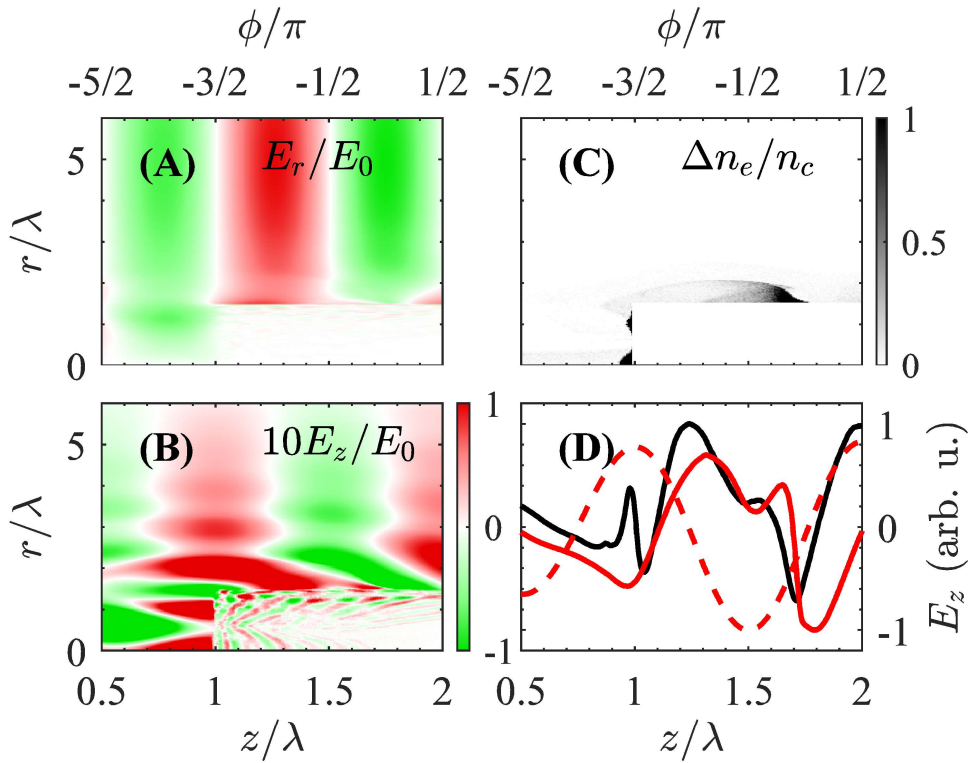

**Fig. 2. Perturbation of the radial and axial electric fields and the electron density.** Snapshots at $t = 1.5\tau$ of: (A) $E_r$, and (B) $E_z$, around the wire target. The electric field components are scaled by the amplitude $E_0 = E_{r0}$. The gray shade in (C) corresponds to the electron density perturbation $\Delta n_e$. The axial electric field in (B) around the wire is strongly modified by the perturbed electron density. (D) Axial field $E_z$ (in arbitrary units) at $r = 1.05 r_0$ and $r = 3 r_0$ (with $r_0$ the initial wire radius), shown by solid-red and dashed-red curves, respectively. The solid-black line represents the near-wire $E_z$ (at $r = 1.05 r_0$) averaged over a laser period, which corresponds to the contribution of the space-charge.

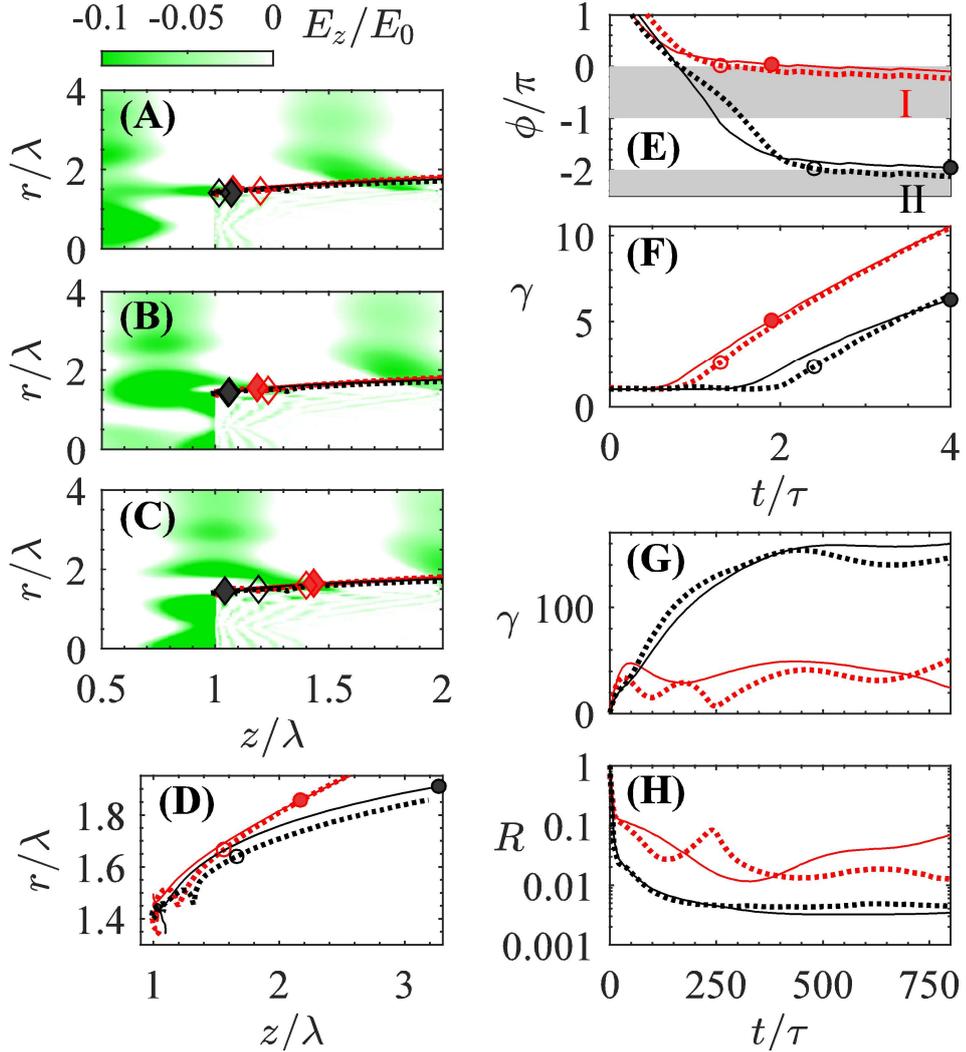

**Fig. 3. Simulation of the ADA process.** Density plots in the $r - z$ plane of the accelerating field ($E_z < 0$) are displayed here as snapshots taken at times: (A) $0.5\tau$, (B) $0.75\tau$, and (C) $\tau$. (D) Trajectories of single electrons which typically originate from the front surface of the wire and stay in the stable phase during acceleration. The same trajectories are also shown in (A)-(C) with their positions at the corresponding times indicated by the diamonds. (E) Variations with time of the phase $\phi$ associated with the electrons considered in (**D**). The gray shaded areas represent acceleration phases I: $[\phi_0 - \pi/2, \phi_0 + \pi/2]$, and II: $[\phi_0 - 5\pi/2, \phi_0 - 3\pi/2]$, while circles indicate the time of electrons entering these phases. (**F**) Evolution in time of the Lorentz factor (or scaled energy) to clearly demonstrate pre-acceleration of the electrons considered in (**D**) and (**E**). Dephasing rates and scaled energies, over longer evolution times, are shown in (**G**) and (**H**).

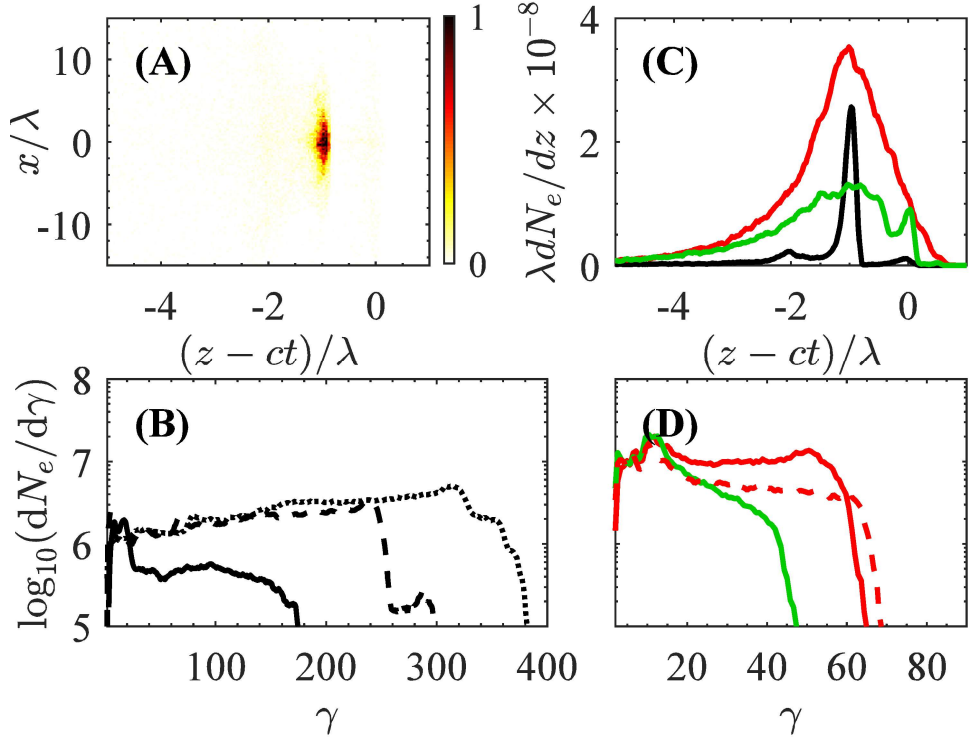

**Fig. 4. 3D-PIC simulation results for energetic electrons driven by lasers of different polarization modes.** (A) Space distribution [$dN_e/(dxdz)$ in arbitrary units] and (B) energy spectra of electrons driven by RP laser pulses. (C) Axial distributions of electrons driven by RP, LP and CP laser pulses with FWHM of 481 as, 5.98 fs and 4.01 fs, and beam charges of 15 pC, 48 pC and 100 pC, respectively. (D) Energy spectra of electrons driven by LP and CP pulses. Parameters in (A) and for the solid lines are same as those in Fig. 3 ($a_0 = 2$ and $n_0 = 20n_c$). The $\gamma$ cut-off values of the solid lines in (B) and (D) are 174, 48, and 64, respectively. Black curves represent results driven by RP pulses at $t = 450\tau$. Green and red curves represent electrons at $t = 150\tau$ driven by LP and CP pulses, respectively. Dashed and dotted lines in (B) correspond to cases with $a_0 = 3$ and 4, respectively. Dashed curve in (D) corresponds to the case with $n_0 = 100n_c$.